\documentclass[copyright,creativecommons]{eptcs}
\providecommand{\event}{}
\providecommand{\volume}{??}
\providecommand{\anno}{20??}
\providecommand{\firstpage}{1}
\usepackage{breakurl}
\usepackage{graphicx}

\title{Diffusion controlled reactions, fluctuation dominated kinetics, and living cell biochemistry}
\author{Zoran Konkoli
\institute{Chalmers University of Technology, Gothenburg, Sweden}
\institute{Department of Microtechnology and Nanoscience - MC2}
\institute{Bionano Systems Laboratory}
\email{zorank@chalmers.se}
}
\def\titlerunning{Diffusion controlled reactions}
\def\authorrunning{Z. Konkoli}
\begin{document}
\maketitle

\begin{abstract}
In recent years considerable portion of the computer science community has focused its attention on understanding living cell biochemistry and efforts to understand such complication reaction environment have spread over wide front, ranging from systems biology approaches, through network analysis (motif identification) towards developing language and simulators for low level biochemical processes. Apart from simulation work, much of the efforts are directed to using mean field equations (equivalent to the equations of classical chemical kinetics) to address various problems (stability, robustness, sensitivity analysis, etc.). Rarely is the use of mean field equations questioned. This review will provide a brief overview of the situations when mean field equations fail and should not be used. These equations can be derived from the theory of diffusion controlled reactions,
and emerge when assumption of perfect mixing is used.
\end{abstract}

\section{Introduction}

The emphasis of this review is on discussing the use of the framework of diffusion controlled reactions to model the living cell biochemistry. No attempt is being made to provide a comprehensive list of references for such a complex field. Instead, selected topics have been carefully chosen to help the non-experts gain the glimpse of the relevant issues. First aim is to introduce general framework for describing living cell biochemistry and give few examples of mathematical models one needs to solve. Second goal is to illustrate that the mean field equations are lowest order approximation of a more complicated set of equations. Final aim is to discuss these equations of motion, the way they are solved, and behavior of solutions.

First, few words about the problem. Chemical reaction kinetics in vivo differs significantly from the one in pipette \cite{ref1}. Geometry of the living cell interior can be quite complicated and there is experimental evidence that the cell is structured in many ways, already starting at the cytoplasm level. For example, for a single cell the total amount of protein content can be as high as 17-30\% by weight which results in extremely structured and crowded space. In addition, the cell interior (roughly $10\mu m$ in diameter) is further partitioned in smaller spaces such as organelles (e.g. mitochondria with 50 nm in diameter), and roughly 50\% of the cell volume is filled by organelles. For typical physiological concentrations of individual proteins of $1nM$ one gets $N_{prot}\sim 1nM(10\mu m)^3\sim 1000$ copies of individual protein. This can result in large spatio-temporal fluctuations of protein number. Also, delivery of proteins can become an issue.

There are two ways to theoretically study such complicated reaction environment, either by performing stochastic simulations or by constructing equations of motion that describe time evolution of averaged quantities (e.g. concentration). The focus will be on the later. It will be argued that any attempt to describe intracellular chemical reactions in terms of mean field equations (that assume perfect mixing) might fail spectacularly. Intracellular dynamics is intrinsically stochastic due to low number of chemicals, exclusion effects are important and, ultimately, many scales interact at the same time with large degree of spatio-temporal organization. To develop theoretical framework that can be used to describe such situation is far from trivial. The theory of diffusion controlled reactions naturally suggests itself in this context. In the following some generic features of diffusion controlled reaction will be discussed with a particular emphasis on mathematical/physical models that are used to describe them. The validity of the truncated equation of motion approach will be critically reviewed, with a particular emphasis on its simplest form, the mean field equations/kineteics.

\section{Diffusion controlled reactions: mathematical model}

Diffusion controlled reactions are ubiquitous in nature. They appear in the matter-antimatter annihilation in the early universe, epidemics spreading and occur frequently at small scales in the living cell. Few references on the topic can be found in \cite{ref2,ref4,ref3,ref5}. Reviews \cite{ref2,ref4} provide gentle introduction to the field. Unfortunately, the field of diffusion controlled reaction is rather technical and paper \cite{ref3} is tutorial of basic techniques one can use. One dimensional systems are reviewed in \cite{ref5}. This work provides rather lengthy account of experimental and theoretical approaches used to describe one dimensional diffusion-controlled reactions.

Perhaps the simplest way to introduce diffusion-controlled reactions is through two stochastic bench mark models: the $A+A\rightarrow P$ or $A+B\rightarrow P$ reaction diffusion models. (For simplicity reasons, both reaction result in same product, though this obviously needs not be the case.) It is assumed that one can clearly separate two time scales in the problem, related to the rates of transport $D$ and the reaction processes $\lambda$.

There are two ways to describe the system, either by using off- or on-lattice models. In this review on-lattice models will be used, thought results obtained for such model hold for off-lattice models as well. For example, assuming that the spacing in the lattice is $h$, continuous models can be obtained by taking the $h\rightarrow 0$ limit, and by re-scaling variables in the appropriate way (e.g. see reference \cite{ref3} for details).

By assumption, particles A and B move on the lattice with jump rates (diffusion constants) $D_A\equiv D$ and $D_B$ respectively.
Further, it is assumed that a pair of particles X and Y at $r$ and $r'$ reacts with rate $\sigma_{X,Y}(r-r')$ where $X,Y\in\{A,B\}$. Here the assumption of translational invariance have been used. For simplicity reasons it will be assumed that $\sigma_{A,A}(r-r')=\lambda\Delta_{r,r'}$ and $\sigma_{A,B}(r-r')=\delta\Delta_{r,r'}$ where $\Delta_{r,r'}$ equals one (zero) for $r=r'$ ($r\ne r'$). More complicated model can be obtained by allowing for finite reaction range but such models are much more technical without conveying anything beyond what short range models already do (within the scope of this review). Please note that in the case of lattice models all rates have the dimension $1/s$. These rates are re-scaled when limit $h\rightarrow 0$ is taken, leading to the parameters that can be directly related to experiments.

\begin{figure}
  \center{\includegraphics[width=8cm]{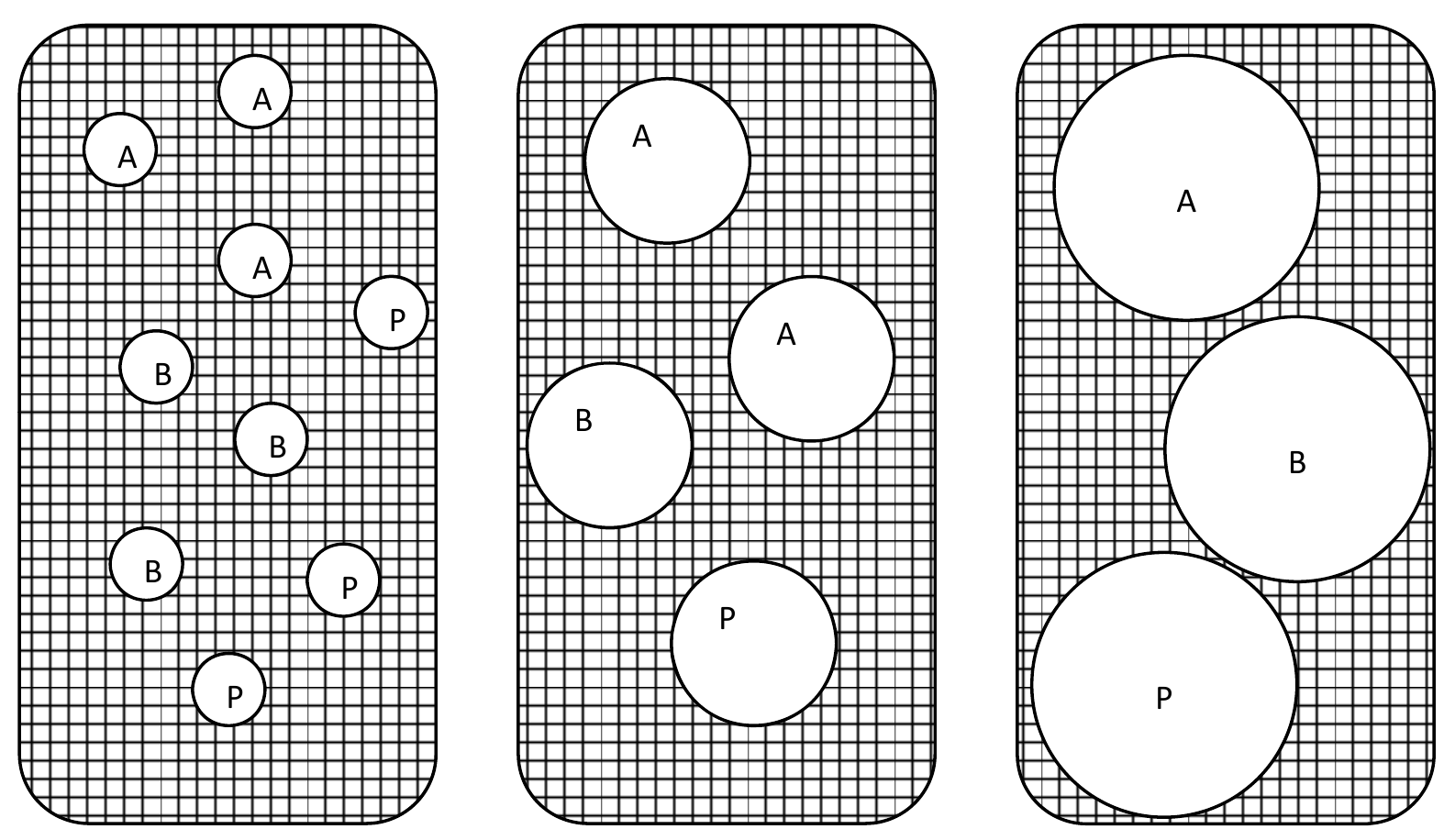}}\\
  \caption{$A+B$ reaction in three different regimes. Panel (a): reactant radius a much smaller than size of the system $L$. In such a case reactants have lot of space to move. Panel (c): crowded situation where $a$ is of the same order of magnitude as $L$. Panel (b) represents intermediate case.}
  \label{fig1}
\end{figure}

Reactants and products are confined in a reaction volume $V\sim L^d$, where $L$ denotes the size of the system and $d$ dimension. (As will be discussed, the dimensionality of the system is the most important parameter that governs validity of mean field equations.) It is assumed that typical size of molecules participating in reactions is $a$. The setup for the A+B reaction is shown in Figure \ref{fig1}. (To visualize A+A system just imagine that all B particles are changed into A.) Particles A and B can be almost anything, molecules, excitons, electron-hole pairs, sellers and buyers on the stock market etc. P denotes reaction product.

Figure \ref{fig1} depicts three distinct regimes. Panel (a) shows a situation when $L\gg a$ and reactants have space to move and do not disturb each other. Such situation is often modeled by assuming that the size of the reaction volume $L$ is infinite. Also it is assumed that products of the reaction $P$ do not exert influence on the reactants leading to a simplification that reactants annihilate without further trace leading to $A+B\rightarrow\emptyset$ (or $A+A\rightarrow\emptyset$). These approximations are very good when $L\gg a$. Panel (c) shows an opposite situation when $a\sim L$ (and still $a<L$). In such a case reactants do not have space to move and one needs to consider exclusion effects. Panel (c) shows intermediate case. In the following the two extreme cases (a) and (b) will be discussed. The living cell harbors reactions with both types of behaviors depending on the relative sizes of reactants and reaction volumes. As time goes on the reaction-diffusion system exhibits variety of behaviors, as discussed in \cite{zk1}.

System constructed so far is stochastic. To study its behavior one could perform stochastic simulations and there are variety of techniques for doing it. However, in the following another approach is followed where equations are constructed that describe time evolution of averaged quantities of interest (various observables such as particle number, correlation functions, variance of particle number etc). In the following section the generic behavior of the diffusion controlled reactions will be discussed. The emphasis will be on intuitive understanding. The technical side of the problem will be presented later.

\section{Qualitative analysis of diffusion controlled reactions in infinite (large) volumes}

\begin{figure}
  \center{\includegraphics[width=8cm]{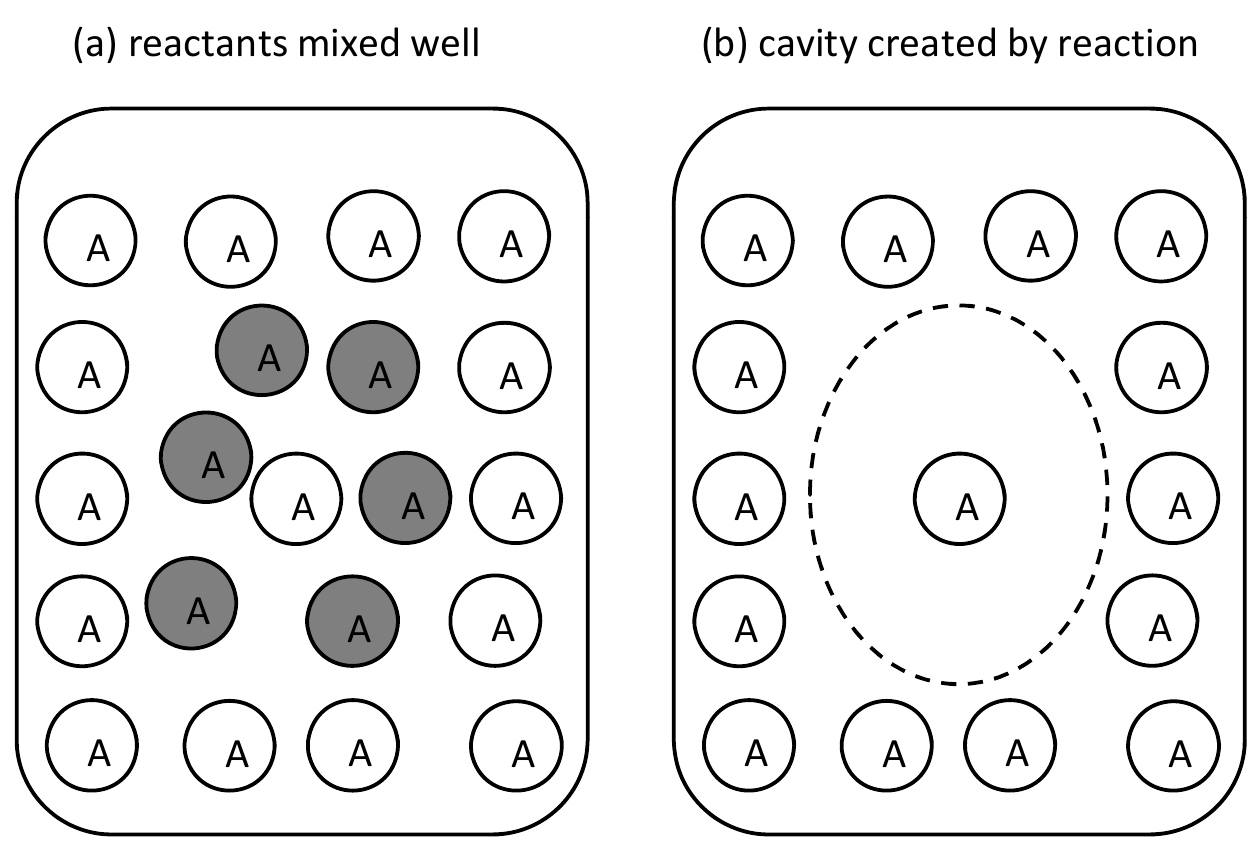}}\\
  \caption{Reactions create spatial fluctuations. Diffusion is slow in filling in the cavity and spatial fluctuations remain leading to the fluctuation dominated kinetics.}
  \label{fig2}
\end{figure}

Figure \ref{fig2} shows a sketch of a snapshot of the $A+A\rightarrow\emptyset$ reaction diffusion system. (Later on real figures from the simulation will be shown.) Interesting phenomena happen when reaction rate $\lambda$ is lot larger than the diffusion rate $D$. Large spatial density fluctuations may develop as time flows. For that particular reason it is far from trivial to predict how particle number $N_A(t)$ or particle density $\rho_A(t)=N_A(t)/V$ vanish in time. The problem at hand is a complex many body problem and the mechanism that governs its behavior is illustrated in the figure.

Even if reactants were mixed well initially, reactions create spatial fluctuations that diffusion cannot smear out. Figure \ref{fig2}, panel (b), shows a cavity that is created when A particles, emphasized by gray shade in panel (a), annihilate. The A particle that is left in the middle of the cavity will not be annihilated unless diffusion fills in the cavity by other A particles. However, if diffusion processes are much slower than the reaction processes $D\ll\lambda$ it takes longer time to fill the cavity. Even if this cavity is filled by particles there will be other cavities that emerge due to the reactions going on. In such a case dynamics of the system is plagued by spatial fluctuations in particle density. Since particles are not mixed well it takes longer time to annihilate all particles.

\begin{figure}
  \center{\includegraphics[width=8cm]{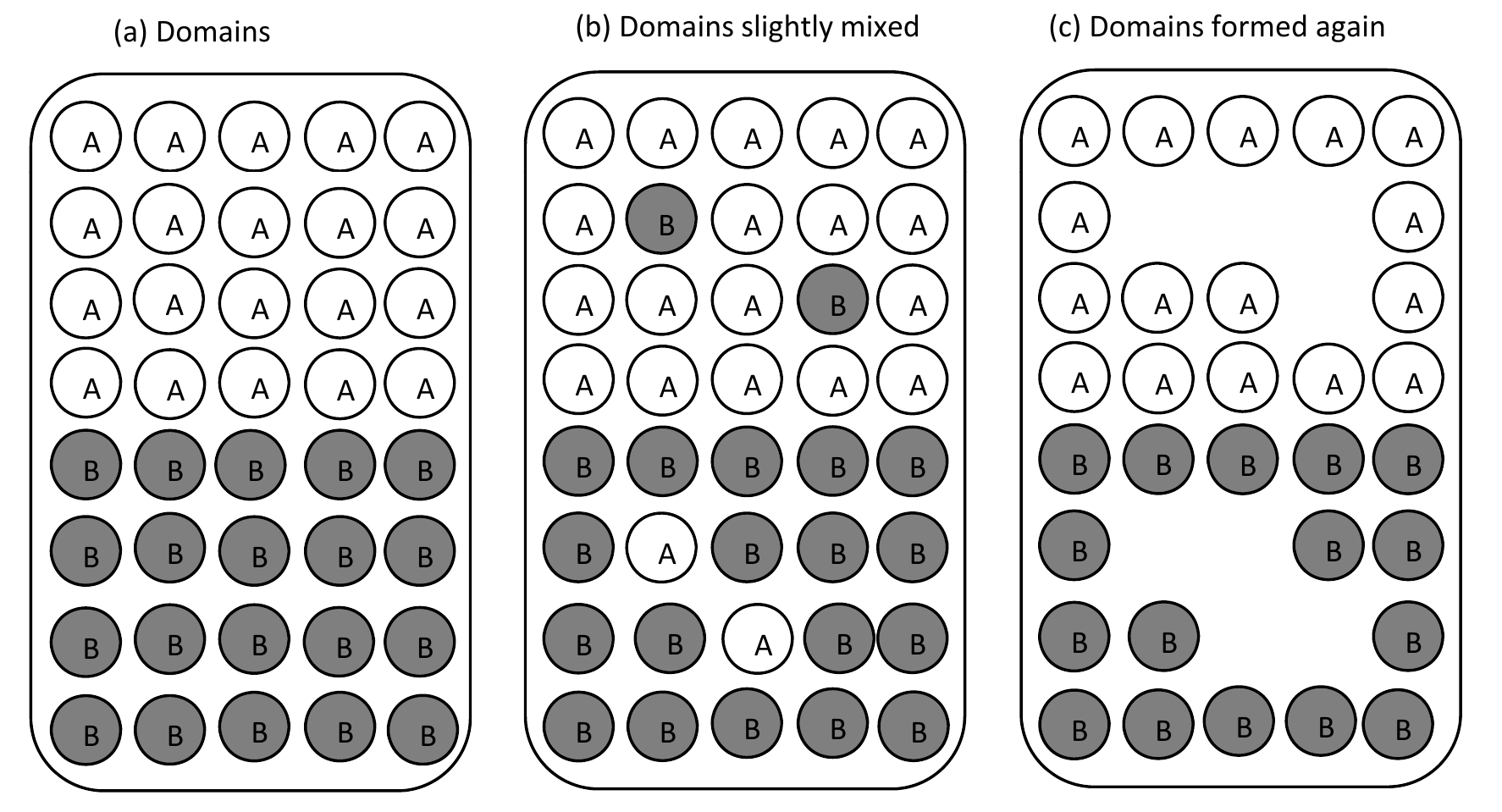}}\\
  \caption{Spontaneous formation of domains for A+B reaction. Panel (a): Assumption is made that domains are formed. Panel (b): domains are mixed by diffusion. Diffusion process should ruin domain structure. Panel (c): Domains get reestablished due to the presence of reactions.}
  \label{fig3}
\end{figure}

The fluctuation dominate regime of A+B reaction looks slightly different from the one for A+A reaction. The behavior of the system differs significantly depending on whether $D_A=D_B$. In the following such situation will be only discussed. Figure \ref{fig3} depicts typical particle distribution when particles are mixed well (e.g. at t=0) and after the reactions proceed for some time. If $D\ll\delta$ domains rich in $A$ or $B$ particles develop.

\begin{figure}
  \center{\includegraphics[width=8cm]{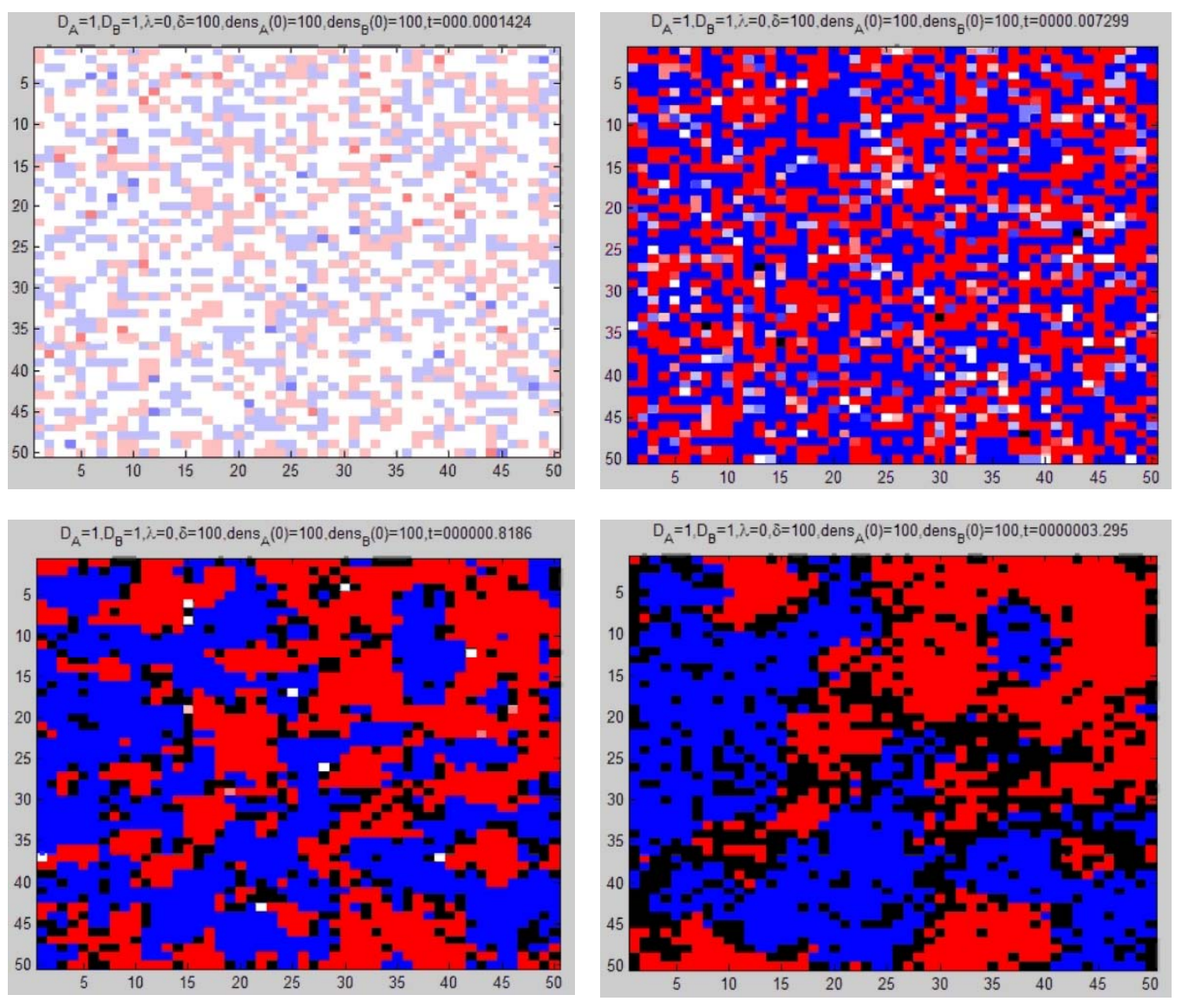}}\\
  \caption{Snapshots of the A+B reaction dynamics in fluctuation dominated regime. Red denotes A particles, B denotes blue particles, and white indicates same number of A and B particles. Intermediate colors correspond to intermediate cases. Black color denotes absence of particles. One can see that domains rich in A and B particles are formed as time goes on.}
  \label{fig4}
\end{figure}

Figure \ref{fig3} shows how reactions lead to spontaneous formation of domains. Even if diffusion tried to mix particles minority species is immediately annihilated. For example, upper part shows a situation when two B molecules (in such a case minority species) diffuses into the domain rich with A particles. B molecules will be immediately annihilated. Same holds for A molecules in the lower part of the panel (b). The mechanism just discussed indeed show up in the snapshots of the particle distribution that originate from stochastic computer simulations. Please see Figure \ref{fig4} for details. The simulation in Figure \ref{fig4} is done using improved minimal process algorithm \cite{ref6} and clearly shows formation of domains. This is a well know result \cite{ref8}.

\section{Mathematical analysis of the diffusion controlled reactions: pair approach}

The type of the kinetics discussed in the previous section is referred to as {\em fluctuation dominated kinetics} since presence of reactions lead to appearance of spatio-temporal fluctuations. Equally often the term {\em anomalous kinetics} is used to describe such systems where in this context "anomalous" indicates deviation from mean field kinetics. The fluctuation dominated kinetics cannot be described using mean field equations and in this section it will be discussed where the problem is.

The problem at hand, i.e. a set of particles moving and reacting on the lattice can be described in terms of the master equation,
\begin{equation}\label{meq}
    \dot  P(c,t) = \sum_{c'} \left[
        R_{c'\rightarrow c} P(c',t) - R_{c\rightarrow c'} P(c,t)
    \right]
\end{equation}
where $P(c,t)$ describes probability that system is in the state $c$ at time $t$. The $R(c\rightarrow c')$ denote transition rates which can be deduced from details of the model described previously. Here and in the following the dot over a symbol denotes the time derivative.

The observables of interest denote average quantities such as densities $\rho(r,t)$ or correlation functions $\rho(r,r',t)$ where $r$, $r'$, etc point to lattice sites. There are couple of ways how to obtain equations for time dependence of observables. For example one can map the problem in Eq.~(\ref{meq}) to field theory and use various field-theoretic techniques to construct equations of motion.~\cite{MG,ref14} An alternative to this approach is to construct equations directly starting from (\ref{meq}), as discussed in \cite{ref3}. In either case calculations are rather technical and will be omitted. Only final form of equations will be stated here. Equations of motion have the same form regardless whether multiple occupancy or single occupancy of lattice sites are allowed (though form of boundary conditions might differ). Work in \cite{ref14} focuses on multiple occupancy situation while work in \cite{ref3} deals with single occupancy situation. For large volumes both approaches give essentially the same results.

To illustrate how the mean field equations emerge it is useful to look at the simplest possible example, the A+A annihilation process. After taking the continuum limit (e.g. $x=hr$, $x'=hr'$, etc) the equations that describe $A+A$ reaction-diffusion system are given by
\begin{equation}\label{aa1}
    \dot \rho(x,t) = D \nabla_x^2 \rho(x,t) - \int dV' \sigma(x-x') \rho(x,x',t)
\end{equation}
and two point density is defined through
\begin{eqnarray}
\dot \rho(x,x',t) &=& D (\nabla_x^2+\nabla_{x'}^2)\rho(x,x',t)
    - \sigma(x-x') \rho(x,x',t) \nonumber \\
   & & - \int dV'' \left[ \sigma(x-x'') + \sigma(x'-x'') \right] \rho(x,x',x'',t)
   \label{aa2}
\end{eqnarray}

Equation (\ref{aa2}) shows that two point density depends on three point density. Such behavior continues ad infinitum. One needs to cut the hierarchy in order to be able to use it. The most common approximation is pair approximation, where book-keeping of pair correlations is only done. This amounts to approximating three point density as \cite{ref3,ref14}
\begin{equation}\label{rho2}
    \rho(x,x',x'',t)\approx \rho(x,t) \rho(x',t) \rho(x'',t)
    \chi(x,x',t) \chi(x',x'',t) \chi(x,x'',t)
\end{equation}
where $\chi(x,x',t)$ denotes pair correlation function defined through $\rho(x,x',t)\equiv \rho(x,t)\rho(x',t)\chi(x,x',t)$. Using the pair approximation and after assuming translational invariance, $\rho(x,t)=\rho(t)$ and $\chi(x,x',t)=\chi(x-x',t)$, results in following equations:
\begin{equation}\label{rho1}
    \dot \rho(t) = - k(t) \rho(t)^2
\end{equation}
where $k(t)=\int dV \sigma(x) \chi(x,t)$ and pair correlation function satisfies
\begin{equation}\label{chi}
    \dot \chi(x,t) = 2D \nabla_x^2 \chi(x,t) - \sigma(x)\chi(x,t)
\end{equation}
with initial condition given by $\chi(x,t=0)=1$, which denotes perfect mixing at t=0.

Equation (\ref{rho1}) is always exact while Eq.~(\ref{chi}) is valid only in the pair approximation. One can see from Eq.~(\ref{rho1}) how mean field result emerges. If $\chi(x,t)=1$, i.e. reactants are mixed all the time, $k(t)=\lambda$ becomes constant and one obtains law of classical chemical kinetics, with a solution $\rho(t) = \rho_0/(1 + \lambda \rho_0 t)$
%
%
and for large times one has $\rho(t)\approx 1/\lambda t \propto t^{-1}$. The -1 is referred to as a mean field exponent.

From figures \ref{fig2}, \ref{fig3}, and \ref{fig4} it is clear that $\chi(x,t)\ne 1$ since diffusion can not mix reactants well. Such behavior can be traced back to the Eq.~(\ref{chi}) where first term describes mixing by diffusion and second term how reactions burn hole in the pair correlation function. Once the equation is solved one can see that for $D\ll\lambda$ and $d\le 2$ pair correlation function never equals one.

To get the correct kinetics rather technical analysis needs to be done for each particular case. Such analysis is beyond the scope of this review. The results are as follows. For A+A reaction the mean field result holds for $d>2$. For $d<2$ the correct result is given by
\begin{equation}\label{aa}
    \rho(t)\approx {\cal A}(d) (Dt)^{-d/2}
\end{equation}
where constant $\cal A$ only depends on the dimensionality of the problem and dependence on the reaction rate $\lambda$  has been lost. $d/2$ is the anomalous kinetics exponent (different from mean field exponent for $d<2$). At $d=2$ one obtains logarithmic corrections, $\rho(t)\propto \ln t/t$.

For the A+B model one has following behavior. To save the space only the case with equal diffusion constants $D_A=D_B=D$ and equal amount of A and B particles at t=0, $\rho_A(x,0)=\rho_B(x,0)=\rho_0$, will be discussed. In such a case mean field equations reduce to ordinary differential equations in time; $\dot\rho_A(t)=\dot\rho_B(t)=-\delta\rho_A(t)\rho_B(t)$. One has the same scaling behavior as in the case of A+A reaction; $\rho_A(t)=\rho_B(t)\approx  1/\delta t$. This is the correct result for $d>4$. However, for $d<4$ one has anomalous kinetics with exponent $d/4$,
\begin{equation}\label{ab}
    \rho_A(t)=\rho_B(t)\approx {\cal B}(\rho_0,d) (Dt)^{-d/4}
\end{equation}
The constant  $\cal B$ depends both on dimensionality and the initial concentration of particles. To see technical details of the calculations that lead to the results in (\ref{aa}) and (\ref{ab}) please see \cite{ref7} and \cite{ref8,ref9,ref10} respectively and references therein. The calculations are rather technical. The A+B has not been solved yet bellow and at $d=2$, though exact derivation of lower and upper bounds exists that confirm critical exponent in (\ref{ab}).

To illustrate importance of fluctuation dominated kinetics when studying living cell biochemistry an particular example will be discussed. The model where A+B, A+B, and B+B reactions occur at the same time was suggested previously \cite{ref11,ref12,ref13,ref14}. The behavior of the system depends very much on the equality of diffusion constants. Assumptions $D_A\ne D_B$ \cite{ref11} and $D_A =D_B$ \cite{ref12,ref13,ref14} stem from different setups and lead to very different behavior of the system.

The $D_A=D_B$ case is a notorious example of how mean field can go wrong. In particular the ABBA model suggested and studied in \cite{ref12,ref13,ref14} shows dramatic failure of mean field approach. The model arises when one wants to include steric effects into the simplest possible way and results in the following set of reactions: A+A, A+B, and B+B with rates $\lambda$, $\delta$, and $\lambda$ respectively. Please note that the A+A and B+B reactions have the same rate constant. Also rate constants are such that $\delta>\lambda$. (The opposite case can be also studied but is not interesting in the context of this review). The mean field equations for such model are given by $\dot\rho_A = -\lambda\rho_A^2 - \delta\rho_A\rho_B$ and $\dot\rho_B = -\lambda\rho_B^2 - \delta\rho_A\rho_B$.
%
%
By mapping the problem to Poincare sphere \cite{ref13} large time behavior is given by
%
%
$\rho_A(t)/\rho_B(t) \propto t^{\delta/\lambda-1}$, when $\rho_A(0)>\rho_B(0)$. The type of molecule that is minority at t=0 simply has to vanish asymptotically. Majority species survives only. However, the careful calculation \cite{ref13,ref14} reveals that the ratio $\rho_A(t)/\rho_B(t)$ saturates to constant for large times and minority species survives. This is an example that the mean field description spectacularly fails to describe.

\section{Diffusion controlled reactions in restricted geometries}

In previous subsections diffusion controlled reactions in the infinite volumes have been discussed. In here, focus will be shifted on discussing diffusion controlled reactions in restricted geometries. A typical situation of interest is depicted in figure \ref{fig1}, panel (c). When size of the reactants is comparable to the size of system extreme crowding conditions arise and in situation like this one cannot neglect presence of product molecules (even if back reaction is absent).

Compared to the infinite volume case diffusion controlled reactions in small (restricted) volumes are relatively unexplored. There are some pioneering efforts in this area, e.g. the work by Khairutdinov and Serpone, Tachiya, or Ramamurthy; see references \cite{ref15,ref16,ref17} for reviews on the subject. In such systems the number of reactants tends to be small and the effects of fluctuations are enhanced. The methods used to study bulk situation (infinite volume), if they indeed work, have to be heavily modified.

The main difference from the case when the volume is finite is that the particle decay is governed by an exponential decay law, instead of the power law found in infinite systems. Such behavior comes from the fact that configuration space becomes strictly discrete. When number of particles becomes low one needs to count them one by one. The average of the number of particles behaves as $n(t)=c_1+c_2 e^{-E_1 t}+ c_3 e^{-E_2 t}+\ldots$. In such a case the transition rate operator entering master equation attains eigenvalues that are well separated. And when time is large enough one has $(E_1-E_i)t\gg 1$ for $i=2,3,4,\ldots$ and first two terms in the expansion for $n(t)$ dominate resulting in the exponential decay. Please note that even the very large system with large number of particles might eventually arrive in the regime where particle/density decay is exponential (e.g. as discussed in \cite{zk1}).

One can naturally wonder about the usefulness of pair approach in the context of reactions in restricted geometries. It was shown that approach can be used but it has severe drawback in small reactions volumes when symmetries or conservation laws exist \cite{zk1}.

\section{Relevance for understanding the living cell biochemistry}

The importance of proper modeling of the living cell biochemistry has been discussed in \cite{ref18} and there are couple of reasons why the framework of diffusion controlled reactions and the fluctuation dominated kinetics are relevant for modeling of the living cell biochemistry. The diffusion is the ubiquitous transport mechanism in the living cell and there are one million reactions happening in the cell per second. In that sense the framework of \emph{diffusion} controlled \emph{reactions} is certainly right choice of the computational platform.

Many important biochemical details are not addressed in the reaction-diffusion framework such as effects related to structured water, chain like structure of molecules, or processes on the time scales much smaller than the reaction times. Nevertheless, the formalism captures the most important aspects of the problem with a clear separation between transport and reaction processes and provides a well defined mathematical formulation of the problem. The structure of the configuration space and transitions among the states are well defined. Calculation of observables is straight forward, thought technically complicated. Also, should a need occur one can perform simulation of the system based on the reaction-diffusion framework.

If one wants to use the diffusion controlled reactions framework couples of issues have to be dealt with. For example, it is not clear whether to use rather technical diffusion-controlled reactions formalism or simplified version of it such as mean field calculations (classical chemical kinetics). In general, one cannot say when the classical chemical kinetics is applicable and when expect appearance of the fluctuation dominated kinetics. There are series of problems where validity of mean field equations is an issue. For example, the stability, sensitivity, or robustness analysis based on mean field equations. The example of the ABBA model discussed in the review shows that even at qualitative level mean field can fail miserably. Thus in general mean field kinetics should be used with caution. It is important to be aware of the risks.

To determine whether mean field equations are valid rather lengthy and technical analysis needs to be done for each new model. Central issue is to identify the critical dimension of the system. For example, $d_c=2$ for A+A reaction in the infinite volume, and $d_c=4$ for A+B reaction for the situation discussed in previous sections. Unfortunately, power counting techniques from field theory do not always work. For example, using power counting one can predict critical dimension for A+A problem but not for A+B.

The value of critical dimension is determining factor that governs validity of mean field equations. In three dimensions (d=3) A+A reaction does not suffer from anomalous kinetics and mean field (classical kinetics) approach can be safely used. However, if one studies reactions on surface (d=2) or line (d=1) one has to be careful. The situation is more alarming or A+B reaction. The A+B reaction is always critical (it always suffers from anomalous kinetics) regardless whether occurring in bulk (d=3), surface (d=2) or line (d=1) since its critical dimension equals $d_c=4$.

Should one use the computer simulations, there are couple of issues to be aware of. If one solely counts the particles in the cell then perfect mixing is assumed. In such a case effects related to spatio-temporal fluctuations are ignored. To go beyond that simplification and account for position of particles one needs to formalize the problem in mathematical terms and naturally ends up using the framework of diffusion controlled reactions. Simulations done at that level account for  most important aspects of the problem. Also, fluctuation dominated kinetics (if it appears) is automatically taken care of. For example, one does not need to worry about validity of the mean field equations. In that respect, there is no need to perform rather technical and time consuming analysis, such as finding the critical dimension of the system. Simulations are very attractive approach to describing living cell.

However, though extremely useful, in silico experiments are heavily dependent on the computer resources, the cpu power, the memory etc. There is certainly an upper limit on the number of particles one can simulate. The equation of motion approach discussed in this review could be interesting alternative in the situation when there are large number of particles in the system and there are not that many particle types. Admittedly, the situation in the living cell is the opposite (large number of particle types in low copy numbers). Nevertheless, equation of motion approach could be useful since it could be adjusted to describe low particle number as shown in \cite{zk1}.

Extremely important issue is whether effects related to fluctuation dominated kinetics appear in the living cell and to what extent. The properties of the diffusion controlled reactions discussed in previous sections might have profound effects on our understanding of the living cell biochemistry. One of the important question is whether the living cell exploits these properties in any way to successfully perform its function. For example, the living cell environment exhibits kinetics in all dimensions. As a rule of thumb, lowering the dimensionality results in increasing deviation from mean field results. Two dimensional reactions at the membrane surface are certainly very different from the one in the bulk. In that sense fluctuation dominated kinetics should be abundant in the living cell. However, careful analysis needs to be done, since living cell is finite.

The problems discussed so far get even more complicated when structure of the cell is taken into account. New issues emerge that need to be addressed. Understanding interplay between geometrical shape that sustains the reactions and topological structure of the pathways is one of the central problems. Ultimately, the question is whether we can understand the shape of the organelles with reference to the set of reactions they sustain. Unfortunately, the role that the geometry plays for cell function is poorly understood, see review papers \cite{bray,ref18} and references therein. In that context diffusion controlled reaction can provide useful tool for that type of analysis. The mathematical platform for studying geometry-reaction interplay framework (GRIP) have been suggested in \cite{ref18,ref19}.

The framework of diffusion controlled reactions is studied extensively in the  statistical physics and chemistry community. However, the study of the fluctuation dominated kinetics has been extensively done within the statistical physics community. Interestingly, in publications that address computational cell biology comparatively little attention has been payed to the effects related to fluctuation dominated kinetics, and in particular validity of mean field equation, though some work exists (e.g. \cite{fdk-lc1,fdk-lc2}). The goal of this review is to point out these facts.


\bibliographystyle{eptcs} 

\end{document}